\documentclass[proof]{pasj00}
\draft

\SetRunningHead{K. Takahashi et al.}
{Similarity between Molecular Loops and Solar Filaments}

\title{Similarity between the Molecular Loops in the Galactic Center
and the Solar Chromospheric Arch Filaments}
\author{Kunio \textsc{Takahashi}
\thanks{Present address is Japan Agency for Marine-Earth Science and Technology, 3173-25 Showa-machi, Kanazawa-ku, Yokohama, Kanagawa 236-0001}}
\affil{Center for Computational Astrophysics (CfCA), National Astronomical Observatory of Japan,\\ Osawa, Mitaka, Tokyo 181-8588}
\email{kutaka@cfca.jp}
\author{Satoshi \textsc{Nozawa}}
\affil{Department of Science, Ibaraki University, Bunkyo 2-1-1 Mito, Ibaraki 310-8512}
\author{Ryoji \textsc{Matsumoto}}
\affil{Department of Physics, Graduate School of Science, Chiba University, \\ 1-33 Yayoi-cho, Inage-ku,
Chiba 263-8522}
\author{Mami \textsc{Machida}
\thanks{Present address is Department of Astrophysics, Nagoya University, Chikusa-ku, Nagoya 464-8602}}
\affil{Division of Theoretical Astronomy, National Astronomical Observatory of Japan, \\ Osawa,
Mitaka, Tokyo 181-8588}
\author{Yasuo \textsc{Fukui}, Natsuko \textsc{Kudo}, Kazufumi \textsc{Torii}, Hiroaki \textsc{Yamamoto},
\and Motosuji \textsc{Fujishita}}
\affil{Department of Astrophysics, Nagoya University, Chikusa-ku, Nagoya 464-8602}
\KeyWords{Galaxy: Magnetic loops, ISM --- Sun: Chromosphere, Magnetic fields, Magnetohydrodynamics}
\Received{2009 February 1}
\Accepted{2009 May 27}
\Published{}

\begin{document}
\maketitle

\begin{abstract}
We carried out two-dimensional magnetohydrodynamic simulations of the Galactic gas disk to show
that the dense loop-like structures discovered by the Galactic center molecular cloud survey
by NANTEN 4 m telescope can be formed by the buoyant rise of magnetic loops due to the Parker
instability. At the initial state, we assumed a gravitationally stratified disk consisting of the
cool layer ($T \sim 10^3$ K), warm layer ($T \sim 10^4$ K), and hot layer ($T \sim 10^5$ K). Simulation
box is a local part of the disk containing the equatorial plane. The gravitational field is
approximated by that of a point mass at the galactic center. The self-gravity, and the effects of the
galactic rotation are ignored. Numerical results indicate that the length of the magnetic loops emerging
from the disk is determined by the scale height of the hot layer ($\sim$ 100 pc at 1 kpc from the Galactic
center). The loop length, velocity gradient along the loops and large velocity dispersions at their foot
points are consistent with the NANTEN observations. We also show that the loops become top-heavy when the
curvature of the loop is sufficiently small, so that the rising loop accumulates the overlying gas faster
than sliding it down along the loop. This mechanism is similar to that of the formation of solar
chromospheric arch filaments. The molecular loops emerge from the low temperature layer just like the dark
filaments observed in the H$\alpha$ image of the emerging flux region of the sun.
\end{abstract}

\section{Introduction}
Spiral galaxies have large-scale mean magnetic fields. It is widely accepted that the magnetic fields
are amplified and maintained by the dynamo mechanism (e.g, Parker 1971). Mathewson and Ford (1970) showed
by observations of optical polarization due to dust grains that the magnetic fields are nearly parallel to
the galactic plane with  wave-like patterns or loop-like structures, suggesting the emergence of magnetic
field via magnetic buoyancy (e.g., Parker 1966; Mouschovias et al. 1974; Blitz and Shu 1980; see also Tajima
\& Shibata 1997). A typical magnetic field strength in the Galactic plane is a few $\mu$ G. The magnetic field
strength increases with the increase in density toward the Galactic center. According to the radio observations
of the Galactic center and its vicinity, the magnetic field strength reaches a few mG in local regions near the
Galactic center (Morris \& Serabyn 1996).

The Galactic gas disk consists of a gas with three phases, namely, cold ($\sim$ 30 K), warm
($\sim 8\times 10^3$ K), and hot ($\sim 10^6$ K) components (Mckee \& Ostriker 1977). The cold component
corresponds to the molecular clouds observed with radio molecular lines (such as CO lines), while the warm
component is observed with the H\emissiontype{I} line, and the hot component is observed with soft X-rays.

The Galactic molecular cloud survey by NANTEN 4 m telescope (Mizuno \& Fukui 2004) found two dense gas
features having a loop-like shape with a length of several hundred pc and width of $\sim$30 pc within
1 kpc from the Galactic center (Fukui et al. 2006). Fukui et al. (2006) discovered large gradients in
the line-of-sight velocity ($\sim$ 30 km s$^{-1}$) along the molecular loops. Moreover, large velocity
dispersion ($\sim 30-50$ km s$^{-1}$) is observed near the foot-points of the loops. Since the loops
have a total mass of $\sim$ 10$^5\MO$, the kinetic energy of a loop is estimated to be $\sim$ 10$^{51}$ erg.

This energy is too large to be explained by a single supernova explosion. Moreover the velocity distribution
is distinct from that of an expanding shell. These features can be explained by the magnetic loops buoyantly
rising due to the Parker instability (Parker 1966).

Matsumoto et al. (1988) carried out two-dimensional (2D) magnetohydrodynamic (MHD) simulations of
the Parker instability. They found that dense regions are formed in the valley of magnetic loops
in the nonlinear stage of the Parker instability and that shock waves are formed at the footpoints
of the rising magnetic loops where the gas infalling along the magnetic field lines collides with
the dense gas near the equatorial plane of the disk. Shibata \& Matsumoto (1991) applied the results
of the MHD simulations to the formation of molecular clouds in the Orion region. Subsequently, Kamaya
et al. (1996) studied the triggering of the Parker instability by supernova explosions. Basu, Mouschovias
\& Paleologou (1997) investigated the effect of the Parker instability on the structure of the interstellar
medium, and Kim et al. (2000) applied it to galactic disks. Machida, Hayashi \& Matsumoto (2000) reported
the results of the three-dimensional (3D) global MHD simulations of the Parker instability in a
differentially-rotating disk. They found that magnetic loops emerging from the disk form a structured
corona above the disk.

Fukui et al. (2006) presented the results of 2D MHD simulations of local volume of the Galactic disk.
They suggested that the downflows can be the origin of the violent motion \footnote{Among the mysterious
features unique to the Galactic center, the violent motion in the central molecular zone and the other
similar broad velocity features whose velocity dispersion is about 15 to 30 km s$^{-1}$ have been highly
enigmatic for the past few decades.} and extensive heating of the molecular gas in the Galactic center.
In their simulations, however, the cold component of the galactic gas that corresponds to molecular clouds
was not taken into account and the mechanism of the formation of dense loop-like structures was not clear.

In the solar atmosphere, arch filaments are observed in the H$\alpha$ images of the chromosphere. Isobe
et al. (2006) found that dense filaments similar to H$\alpha$ arch filaments are formed in the emerging
flux. These filamentary structures are cool and dense above the chromosphere. This configuration in which
the dense regions exist around the top of the emerging loops is a typical feature of the emerging loops and
is a good approximation to what will be observed as dark features in the H$\alpha$ image. They clarified the
reason why the emerging loop becomes top-heavy on the basis of the results of 3D and 2D MHD simulations of
the emergence of the magnetic loops from the convection zone, through the chromosphere, to the corona. The
dense gas accumulated around the top of the rising loops fragments into filaments by the Rayleigh-Taylor
instability, and slides down along the magnetic field lines. These filaments correspond to the arch filaments
observed in the emerging flux regions of the sun.

In this paper, we present the results of 2D MHD simulations of the Galactic gas disk. We will consider a local volume
of the Galactic gas disk that consists of a low temperature layer (the cold component of the Galactic gas disk), warm
layer and hot layer through 2D MHD simulations. In section 2, we will present basic equations and the numerical model
used in this study. The numerical results will be reported in section 3 and section 4 is devoted for discussion.
Summary will be given in section 5.

\section{Numerical Model}
\subsection{Assumptions and Basic Equations}
We adopt the local Cartesian coordinates $(x,y,z)$ at radius $r_0$ from the Galactic center (see Figure 1).
The $x$-direction is taken to be the azimuthal direction, and the $z$-direction is parallel to the rotational axis of the
Galactic disk. The partial derivatives of the background medium with respect to the radial direction $y$ are neglected
(i.e., local approximation). The assumptions, basic equations and initial conditions are similar to those in Matsumoto et al.
(1988). We will assume the following: (1) the medium is an ideal gas with the specific heat ratio $\gamma=1.05$, (2) the
viscosity and resistivity are neglected, (3) the effects of the rotation of the disk and self-gravity are neglected.

The basic equations are:
\begin{eqnarray}
\frac{\partial\rho}{\partial t}+({\boldsymbol v}\cdot\nabla)\rho
&=&-\rho\nabla\cdot{\boldsymbol v}, \\
\frac{\partial{\boldsymbol v}}{\partial t}+({\boldsymbol v}\cdot\nabla){\boldsymbol v}
&=&-\frac{1}{\rho}\nabla P+\frac{1}{4\pi\rho}(\nabla\times{\boldsymbol B})
\times{\boldsymbol B}+\boldsymbol{g}, \\
\frac{\partial P}{\partial t}+({\boldsymbol v}\cdot\nabla)P
&=&-\gamma P(\nabla\cdot{\boldsymbol v}),
\end{eqnarray}
\noindent and
\begin{eqnarray}
\frac{\partial {\boldsymbol B}}{\partial t}
=\nabla\times({\boldsymbol v}\times{\boldsymbol B}).
\end{eqnarray}
Here, $\boldsymbol{g}=(0,0,g(z))$ is the gravitational acceleration assumed to be
\begin{eqnarray}
g(z)=-GMz/(r_0^2+z^2)^{3/2},
\end{eqnarray}
where $G$ is the gravitational constant. This vertical component of the gravitational acceleration is produced by the
gravitational potential by a point mass and gives a simple but exact model for the accretion disk rotating around the
point mass. In the case of galaxies, this is not an exact model, but at least the region of the equator can be
approximated by this model, because the gravitational acceleration near the disk in galaxies is approximately
proportional to the height from the equatorial plane of the disk. The radial component of gravity is assumed to be equal
to the centrifugal force due to the rotation of the disk. The effects of the rotation will be discussed in section 4.4.

Equations (1)--(4) are rendered dimension less by using normalizing constants $r_0$, $C_\mathrm{S0}$ and $\rho_0$, where
$C_\mathrm{S0}$ is the sound speed in the mid-temperature region and $\rho_0$ is the unperturbed density at the equatorial
plane. When the numerical results are compared with observations, we use the units of length, velocity, and time in the
simulation to be $r_0=1$ kpc, $C_\mathrm{S0}=  18$ km s$^{-1}$, and $t_0 = r_0/C_\mathrm{S0} = 5.6 \times 10^7$ yr, respectively.
The unit temperature is $T_0=\mu C_\mathrm{S0}^2/(\gamma R) = 2 \times 10^4$ K, where $\mu$ and $R$ are the mean molecular weight
and gas constant, respectively. The normalization units are summarized in table 1.

We also introduce a non-dimensional parameter $\varepsilon=\gamma(V_\mathrm{K}^2/C_\mathrm{S0}^2$), where
$V_\mathrm{K}=(GM/r_0)^{1/2}$ is the Keplerian velocity at radius $r_0$. In this simulation, we adopt $\varepsilon=130$ and
$V_\mathrm{K}=200$ km s$^{-1}$. For this parameter, the sound speed in the low-temperature region is about 5.6 km s$^{-1}$
whose value is close to the observed velocity dispersion ($5-9$ km s$^{-1}$) of the main gas components (Boulares \& Cox 1990).
However, this sound speed is larger than that of the molecular gas.

\subsection{Initial State}
The initial state is assumed to be in magnetohydrostatic equilibrium. The gas layer is initially composed of three
layers: the cool equatorial layer ($T=T_\mathrm{c}, |z|<z_1$), the warm (mid-temperature) layer
($T=T_\mathrm{m}, z_1\le|z|\le z_2$), and the hot galactic halo ($T=T_\mathrm{h}, |z|>z_2$). The initial
distribution of temperature is assumed to be
\begin{eqnarray}
T(z)=T_\mathrm{c}&+&(T_\mathrm{m}-T_\mathrm{c})
\left\{\frac{1}{2}\left[\tanh\left(\frac{|z|-z_1}{w_\mathrm{t}}\right)+1\right]\right\}
\nonumber \\
&+&(T_\mathrm{h}-T_\mathrm{m})
\left\{\frac{1}{2}\left[\tanh\left(\frac{|z|-z_2}{w_\mathrm{t}}\right)+1\right]\right\}.
\end{eqnarray}
In this study, we take $T_\mathrm{c}=0.1T_0$, $T_\mathrm{m}=T_0$, $T_\mathrm{h}=10T_0$, $w_\mathrm{t}=0.015r_0$, $z_1=0.12r_0$ and
$z_2=0.24r_0$. Although these values are not realistic for the Galactic gas disk, they would be acceptable for our first attempt to study the effects
of the cold component of the galactic gas and the mechanism of the formation of dense loop-like structures. Numerical results do not depend much on
the temperature of the hot galactic halo (Kamaya et al. 1997). Further discussion on the dependence of numerical results on the disk temperature will
be given in section 4.5.

We assume that the magnetic field is initially parallel to the equatorial plane;
$\boldsymbol{B}=(B_\mathrm{x}(z),0,0)$, and is localized in the cool equatorial region ($|z|<z_1$). The magnetic field strength
is determined by introducing the plasma beta (the ratio of gas pressure to magnetic pressure, hereafter denoted by $\beta$) as
\begin{eqnarray}
B_x(z)=\sqrt{\frac{8\pi P(z)}{\beta(z)}},
\end{eqnarray}
where
\begin{eqnarray}
\frac{1}{\beta(z)} = \frac{1}{\beta_0}\left\{1-\frac{1}{2}\left[\tanh\left(\frac{|z|-z_\mathrm{f}}{w_\mathrm{t}}\right)
+1\right]\right\}.
\end{eqnarray}
Here, $\beta_0$ is the plasma $\beta$ at the galactic plane, and $z_\mathrm{f}$ is the half thickness of the magnetic flux
sheet. The initial density and gas pressure distributions are calculated numerically by solving the equation of
magnetohydrostatic equilibrium:
\begin{eqnarray}
\frac{d}{dz}\left[P(z)+\frac{B_x^2(z)}{8\pi}\right]+\rho(z)g(z)=0.
\end{eqnarray}
We have studied the evolution for five models, whose parameters are summarized in Table 2. The distributions of the initial
temperature $T$, density $\rho$, gas pressure $P$, and magnetic pressure $B_x^2/(8\pi)$ are shown in Figure 2a. Figure 2b
shows the profile of the initial local pressure scale height $\Lambda(z)=C_\mathrm{S}(z)^2/(\gamma g(z))$ and the pressure
scale height including the magnetic field $\Lambda_\mathrm{B}(z)=(1+\beta(z)^{-1})\Lambda(z)$ for model B ($\beta_0=1$),
which we call the fiducial model hereafter.

\subsection{Stability of the Equilibrium Model}
The equilibrium state we described in section 2.2 is unstable against the Parker instability. The Parker instability in an
isothermal gas has the linear maximum growth rate at a finite wavelength $\lambda_\mathrm{Parker}\sim 10\Lambda-20\Lambda$,
where $\Lambda$ is the scale height (Parker 1966). This is because small-wavelength modes are stabilized by the magnetic
tension force. We analyzed the linear stability of the initial model with a normal mode method similar to that of
Horiuchi et al. (1988). The linearized equations are the same as those in Horiuchi et al. (1988) and Kamaya et al. (1997).
We consider the growth of a small perturbation that has a functional form $\delta W \propto \exp(i\omega t + i k_x x +i k_y y)$,
where $W$ is a physical quantity $(\rho,P,v_z,B_x)$ and $\delta W$ is its perturbation. The eigenvalues ($\omega$) and the
eigenfunctions are calculated numerically.

Figure 3a shows the linear growth rate ($i\omega$) of the fundamental mode (Horiuchi et al. 1988) of the Parker instability
as a function of the horizontal wave number $k_x$ for five cases, $\beta_0 = 0.5, 1, 2, 4$ and 10 when
$z_\mathrm{f}/r_0 = 0.08$ and $k_y = 0$. Figure 3b shows that the maximum growth rate is inversely proportional to the
square root of $\beta$ ($i\omega \propto \beta_0^{-1/2}$). This result is consistent with that of the linear analysis of the
Parker instability in uniform gravitational fields (Parker 1966). The most unstable wavelength ($\lambda_\mathrm{max}$) is
$\lambda_\mathrm{max}\simeq (\pi/8) r_0$ for model B ($\beta_0=1$). It is noted from Figure 2b and Figure 3  that $\lambda_\mathrm{max}$
is nearly 10 times the local pressure scale height of the mid-temperature region in $0.12 \le z/r_0 \le 0.24$
($\lambda_\mathrm{max}\sim10\Lambda\sim 400$ pc) and is much larger than that of the low-temperature cool region around $z/r_0 \sim 0.1$.

\subsection{Boundary Conditions and Numerical Method}
We assumed free boundaries at $z=Z_\mathrm{min}$ and $z=Z_\mathrm{max}$ such that waves transmit freely by setting
$z$-derivatives of all the variables vanish, and imposed periodic boundary conditions at $x=X_\mathrm{min}$ and
$x=X_\mathrm{max}$. 

In order to trigger the Parker instability, small velocity perturbations of the form
\begin{eqnarray}
v_z= A C_\mathrm{S0} \cos \left(\frac{2\pi x}{\lambda_x} \right),
\end{eqnarray}
are given initially within a finite horizontal domain ($ |x| \le \lambda_x/2$) on the magnetic flux sheet, where
the perturbation wavelength $\lambda_x$ is close to that of the most unstable wavelength for the Parker instability.
Here $A$ (= 0.01) is the maximum value of $v_z/C_\mathrm{S0}$ in the initial perturbation.

The numerical scheme we used is the Rational CIP (Cubic interpolated profile) method (Yabe \& Aoki 1991; Xiao, Yabe \& Ito 1996)
combined with the MOC-CT method (Evans \& Hawley 1988; Stone \& Norman 1992). The magnetic induction equation was solved by the
MOC-CT and the other equations were solved by the CIP (Kudoh, Matsumoto \& Shibata 1998, 1999).

The size of the simulation box is $(X_\mathrm{max}-X_\mathrm{min}, Z_\mathrm{max}-Z_\mathrm{min}) =
(4r_0, 4r_0) = (19H, 19H)$, where $H$ is the scale height at the point where the gravity is maximum
($H=C_\mathrm{S}^2/(\gamma g_\mathrm{max})=0.21r_0$). The grid sizes are
$\Delta x = 2.5\times 10^{-3}r_0$ for $|x|\leq 0.8r_0$, $\Delta z = 2.5\times 10^{-3} r_0$ for $|z|\leq 0.8r_0$ and they slowly
increase for $|x| > 0.8r_0$, or $|z| > 0.8r_0$ by an increment of 5\% at each grid (e.g, $|\Delta x_{i+1}|=1.05|\Delta x_i|$).
The number of grid points is $(N_x, N_z)=(472, 472)$.

\section{Numerical Results}
Figure 4 shows the time evolution of the fiducial model. Solid curves depict magnetic field lines. Color shows the density
distribution. Arrows show velocity vectors. The overall evolution agrees with that of Matsumoto et al. (1988). That is to say,
as the instability grows, the magnetic field lines bend across the equatorial plane. As the gas slides down along the
undulating magnetic field lines, the rarefied regions buoyantly rise, and form magnetic loops in the later (nonlinear) phase.
In the valleys of the magnetic loops, dense spur-like structures are created almost perpendicular to the galactic plane. At the top
of the emerging loops, dense shell-like structures are formed (Figure 5a). These structures were not recognized in previous simulations
(e.g., Matsumoto et al. 1988) partly because they assumed isothermal atmosphere without a steep density gradient at the disk-halo
interface, and partly because of the lower numerical resolution. Since the downflow speed exceeds the local sound speed, strong shock
waves are formed at the magnetic loop footpoints. Figure 5b shows that the shock waves heat the cool gas around $z/r_0 \sim 0.1$.

At the final phase ($t/t_0=1.6$), the expansion of the magnetic loops is stalled because the driving forces diminish at the top of the
loop. Figure 6 shows the vertical distribution of the magnetic pressure, gas pressure, $\beta$, and density at the midpoint of the
emerging loop ($x/r_0=0.25$). From this figure, it is clear that the magnetic pressure and gas pressure gradients nearly balance at
the top of the emerging loop. This magneto-hydrostatic state is attained because the Parker instability is stabilized in the hot layer
where the local pressure scale height becomes larger than the length of the magnetic loop.

Figure 7 shows the snapshots of the density distribution for all models ($\beta_0=0.5, 1, 2, 4$ and 10) at the stage when the top of
the magnetic loops enters the hot region ($z/r_0 > 0.24$). The time scale for the loop emergence is shorter for lower $\beta$,
consistent with the results of the linear analysis presented in section 2.3. Figure 8 shows the vertical distributions of the density
at the midpoint of the emerging loop. The density increases with height at the top of the magnetic loops in all models.
In the model with $\beta_0=10$, the emerging loop stalls at a lower height ($z/r_0<0.3$) because the released magnetic energy is small.

\section{Discussion}
\subsection{Formation of Dense Loop Structure}
Let us discuss why the emerging loop becomes top-heavy. Figure 9 shows the magnetic field lines at $t/t_0=1.4$ (solid curves) and
$t/t_0=0.9$ (broken curves) for the fiducial model (model B). These field lines are iso-contours of the $y$-component of the vector
potential. The magnetic field lines comoving with the plasma in ideal MHD can be identified by the value of the vector potential.
Since the frozen-in condition is assumed and the numerical diffusion is negligibly small, the field lines at $t/t_0=0.9$ (broken curves)
have moved with gas to the corresponding field lines at $t/t_0=1.4$ (solid curves). As the magnetic loops rise, they can accumulate the
gas above the loop when the loop top is flat. Figure 9 shows that the loop shape at $t/t_0=0.9$ is favorable for the mass accumulation
around the loop top. At $t/t_0=1.4$, since the curvature increases, the mass accumulated around the loop top slides down along the
magnetic field lines. Flat top loops are formed when the loop length is much longer than the local pressure scale height. In our model
atmosphere, since the local pressure scale height decreases with height in $0 < z/r_0 < 0.1$ and increases with height in
$0.1 < z/r_0 < 0.25$ (see Figure 2b), the loop top tends to be flat when its height is $z/r_0 \sim 0.1$. The dense shell around the loop
top is formed when $0.1 < z/r_0 < 0.2$, and the mass drains as the loop rises.

In order to quantitatively evaluate the density enhancement at the top of the emerging loop, we adopt the same method of analysis
as Isobe et al. (2006). The velocity vector $\boldsymbol{v}$ can be divided into two components:
\begin{eqnarray}
\boldsymbol{v}=\boldsymbol{v}_\perp+\boldsymbol{v}_\parallel,
\end{eqnarray}
where $\boldsymbol{v}_\perp$ and
\begin{eqnarray}
\boldsymbol{v}_\parallel=\frac{\boldsymbol{v}\cdot\boldsymbol{B}}{|\boldsymbol{B}^2|}\boldsymbol{B},
\end{eqnarray}
are the velocity components perpendicular and parallel to the magnetic field, respectively.

Figure 10 shows the distribution of $\boldsymbol{v}$ and $\nabla\cdot\boldsymbol{v}$ (Figure 10a,b),
$\boldsymbol{v}_\perp$ and $\nabla\cdot\boldsymbol{v}_\perp$ (Figure 10c,d), and $\boldsymbol{v}_\parallel$ and
$\nabla\cdot\boldsymbol{v}_\parallel$ (Figure 10e,f) at $t/t_0=1.2$ and $t/t_0=1.3$ for the fiducial model. The distributions of
$\nabla\cdot\boldsymbol{v}_\parallel$ and $\nabla\cdot\boldsymbol{v}_\perp$ show that the density inside
the emerging magnetic loops keeps decreasing. However, $\nabla\cdot\boldsymbol{v}$ and $\nabla\cdot\boldsymbol{v}_\perp$
at the top of the emerging loop are negative. Therefore, at least for the present parameters, it can be concluded that the density
increases in the top region of the emerging loop. Figure 11 shows the distribution of the density near the loop top at $t/t_0=1.1$
and $t/t_0=1.2$ for the fiducial model. The clump of gas near the loop top survived due to the small curvature in the low-temperature
layer and is later compressed in the high-temperature layer.

\subsection{Comparison with Observations}
Fukui et al. (2006) found two dense gas features having a loop-like shape with a length of several hundred pc within $\sim$ 1 kpc from
the Galactic center. Figure 3a shows that the most unstable wavelength is likely to be 350 -- 500 pc. Our numerical simulation produced
the magnetic loops whose wavelength is $\sim 400$ pc and their height is $\sim 350$ pc. In the observed molecular loops, the line-of-sight
velocity along the loop changes linearly with the arc-length and has large gradients ($\sim$ 30 km s$^{-1}$). Figure 12 shows the
velocity components along the outermost magnetic field line at $t/t_0=1.4$ for the fiducial model. The gradient of the downflow velocity
given in Figure 12c is $\sim 20-30$ km s$^{-1}$. Since the downflow speed exceeds the local sound speed, strong shock waves are formed at
the magnetic loop footpoints. Such shocks can be the origin of the observed large velocity dispersions near the footpoints of Galactic
center molecular loops.

\subsection{Estimation of the Kinetic Energy of the Emerging Loop}
Fukui et al. (2006) estimated that the kinetic energy of the molecular loop is $\sim 10^{51}$ erg. In our simulation, the kinetic energy
flux, $F_\mathrm{k}$, carried by the downflow is
\begin{eqnarray}
F_\mathrm{k} \simeq \frac{1}{2}\rho_\mathrm{df} v_\mathrm{df}^3 \sim 10^{-5}~\mathrm{erg}~\mathrm{cm}^{-2}~\mathrm{s}^{-1},
\end{eqnarray}
where $\rho_\mathrm{df}$ is assumed to be $10^{-24}$ g cm$^{-3}$ and $v_\mathrm{df}$ is found to be $\sim$ 30 km s$^{-1}$ according to
the numerical result for the fiducial model. The numerical results indicate that the loop width in $z$-direction is about 30 pc
($\sim$ 10$^{20}$ cm) and the downflow continues for about $10^{15}$ s $\sim$ 3 $\times$ 10$^{7}$ yr. When the loop thickness in
$y$-direction is 100 pc, the total kinetic energy of the downflow toward the  both footpoints is $\sim 10^{51}$ erg. This energy is
comparable to that estimated from the observation.

\subsection{Effects of Cooling and Rotation}
In this study, we assumed that the gas layer at the initial state is composed of the cool equatorial layer, warm layer, and hot layers.
We assumed adiabatic gas but in the interstellar gas, cooling and heating play essential roles in the formation of molecular clouds
(Field 1965). Kosi{\'n}ski \& Hanasz (2006, 2007) investigated the Parker instability coupled with thermal processes (cooling and heating).
They found that the Parker instability can trigger the thermal instability which form dense clouds in the valleys of the magnetic loops.
In subsequent papers, we would like to include the gas cooling and heating effects.

We should consider the effect of the rotation of the disk to construct a more realistic model. In rotating disks, the Parker instability
is slightly suppressed by Coriolis forces (Chou et al. 1997). Hanasz et al. (2002) presented the results of resistive 3D MHD simulations
of a local part of the disk including the contribution of the Coriolis force. They demonstrated that the Parker instability, the twisting of
the loops, and magnetic reconnection lead to the formation of helically twisted magnetic flux tubes. It is worth noting other 3D effects such
as the growth of the interchange mode (e.g., Nozawa 2005). Isobe et al. (2006) showed that dense shells at the top of magnetic loops fragment
into filaments due to the growth of the Rayleigh-Taylor instability. In order to include this effect, we need to carry out 3D MHD simulations. 

In Galactic gas disks, magneto-rotational instability (MRI; Balbus \& Hawley 1991) grows and drives magnetic turbulence inside the disk.
In the nonlinear stage, MRI drives magnetic turbulence inside the disk. The effects of stochastic magnetic fields on the growth of the Parker
instability was reported by Parker \& Jokipii (2000). The results of global 3D MHD simulations of the galactic center gas disk are reported by
Machida et al. (2009). 

\subsection{Dependence on the initial disk temperature}
We studied the dependence of the numerical results on the initial disk temperature. Figure 13 shows the  density distribution at the stage
when the top of the magnetic loops reaches $Z/r_0 \sim 0.35$ for models with lower initial disk temperature ($T_\mathrm{c}=0.05T_0$) and
higher initial disk temperature ($T_\mathrm{c}=0.2T_0$). Other paremeters are the same as the fiducial model (model B). Numerical results indicate
that the equatorial dense region at this stage is thin for the cool disk ($T_\mathrm{c}=0.05T_0$) and thick for the warm disk
($T_\mathrm{c}=0.2T_0$). Although the length of the loops near the equator is smaller for cooler disk, the length of the loops emerging in the halo
is almost the same ($\sim 400$ pc). This is because the  most unstable wavelength of the Parker instability is determined by the local pressure scale
height.

\subsection{Other Effects}
Finally, we briefly mention about other effects such as cosmic rays, self-gravity and the presence of a spiral arm of the Galactic gas disk.
In Galactic disks, since the cosmic ray pressure is comparable to the gas pressure, cosmic rays enhance the growth rate of the Parker instability
(e.g., Parker 1966, Hanasz \& Lesch 2000, 2003; Kuwabara, Nakamura \& Ko 2004). The effects of the cosmic rays may be more important in the
Galactic center where strong activities can produce high energy particles.

In this paper, we neglected the self-gravity of the gas. When the surface density of the gas disk is large enough, Parker-Jeans instability grows
(e.g., Elmegreen 1982; Nakamura, Hanawa \& Nakano 1991; Kim, Ostriker \& Stone 2002; Lee et al. 2004). The Jeans instability has a larger growth
rate than does the Parker instability when either the magnetic field is weak or the wavelength of the perturbation is long.

Franco et al. (2002) showed by 3D MHD simulations that the Parker instability creates massive clouds inside the spiral arm of the Galactic gas disk
and that the dense gas accumulated around the equatorial plane forms  corrugated structure. The distribution of H\emissiontype{I} gas and dust
below the molecular loops found in the Galactic center (Torii et al. 2009) is consistent with the simulation by Franco et al. (2002). The density
distribution in the nonlinear stage of our simulation also shows such corrugated structures.

\section{Summary}
In this paper, we have carried out 2D MHD simulations of the Galactic center gas disk consisting of the low-temperature layer and the
mid-temperature layer with the overlying hot halo. We found that the numerical results reproduce basic features in the Galactic-center
molecular loops as observed with NANTEN, such as the loop length, the velocity gradient along the loops, and large velocity dispersions
at the footpoints of the loops.

We also discussed the reason why the top of the emerging loop becomes over-dense. This is because the effective gravity along the
magnetic field lines decreases when the curvature radius of the magnetic field lines increases in the low-temperature layer. The gas
survived near the loop top due to its small curvature is compressed when the loop top enters the high-temperature layer.

Finally, we suggest that the Galactic center molecular loops are analogous to the arch filaments in the solar chromosphere. The
molecular loops emerge from the low temperature layer just like the dark filaments observed in H$\alpha$ images of the emerging
flux region of the sun.\\

We are grateful to T. Sakurai, T. Kudoh and D. Shiota for useful comments and discussion. Numerical computations were carried out
on the general-purpose PC farm at Center for Computational Astrophysics, CfCA, of National Astronomical Observatory of Japan (P.I. KT). 
This work is financially supported in part by a Grant-in- Aid for Scientific Research (KAKENHI) from JSPS (No. 20244014). This work is
also carried out by the joint research program of the Solar-Terrestrial Environment Laboratory, Nagoya University.


\begin{table}[htbp]
\begin{center}
\caption{Units for Normalization}
\begin{tabular}{cccc}
\hline
\hline
Physical Quantity & Symbol & Normalization Unit & Physical Value \\
\hline
Length         & $x,z$  & $r_0$                                & 1 kpc                              \\
Density        & $\rho$ & $\rho_0$                             & $10^{-22}$ g cm$^{-3}$             \\
Velocity       & $v$    & $C_\mathrm{S0}$                      & 18 km s$^{-1}$                     \\
Pressure       & $P$    & $\rho_0 C_\mathrm{S0}^2$             & $3 \times 10^{-10}$ dyne cm$^{-2}$ \\
Magnetic Field & $B$    & $(\rho_0 C_\mathrm{S0}^2)^{1/2}$     & 17$\mu$ G                          \\
Temperature    & $T$    & $\mu C_\mathrm{S0}^2/(\gamma R)$     & 2 $\times$ 10$^4$ K                \\
Time           & $t$    & $r_0/C_\mathrm{S0}$                  & 5.6 $\times 10^7$ yr               \\
\hline
\end{tabular}
\end{center}
\end{table}

\begin{table}[htbp]
\begin{center}
\caption{List of models}
\begin{tabular}{cccccc}
\hline
\hline
Model & $r_0$~[kpc] & $z_1/r_0$ & $z_2/r_0$ & $z_\mathrm{f}/r_0$ & $\beta_0$ \\
\hline
A           & 1 & 0.12 & 0.24  & 0.08  & 0.5 \\
B$^\mathrm{a}$ & 1 & 0.12 & 0.24  & 0.08  & 1   \\
C           & 1 & 0.12 & 0.24  & 0.08  & 2   \\
D           & 1 & 0.12 & 0.24  & 0.08  & 4   \\
E           & 1 & 0.12 & 0.24  & 0.08  & 10  \\
\hline
\end{tabular}
\end{center}
\begin{center}
$^\mathrm{a}$Fiducial model
\end{center}
\end{table}

\begin{figure}
\begin{center}
\FigureFile(100mm,100mm){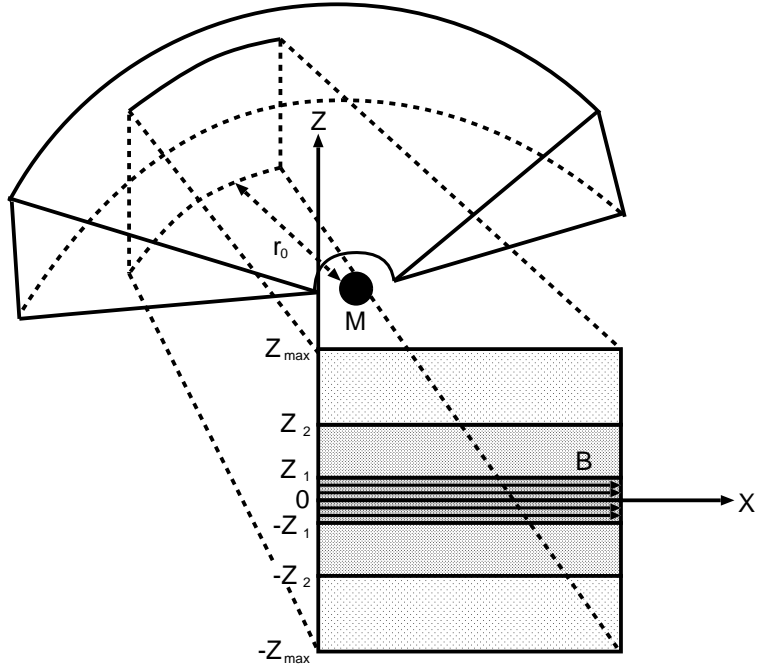}
\end{center}
\caption{Schematic picture of the simulation model and simulation box.}
\end{figure}

\begin{figure}
\begin{center}
\FigureFile(160mm,160mm){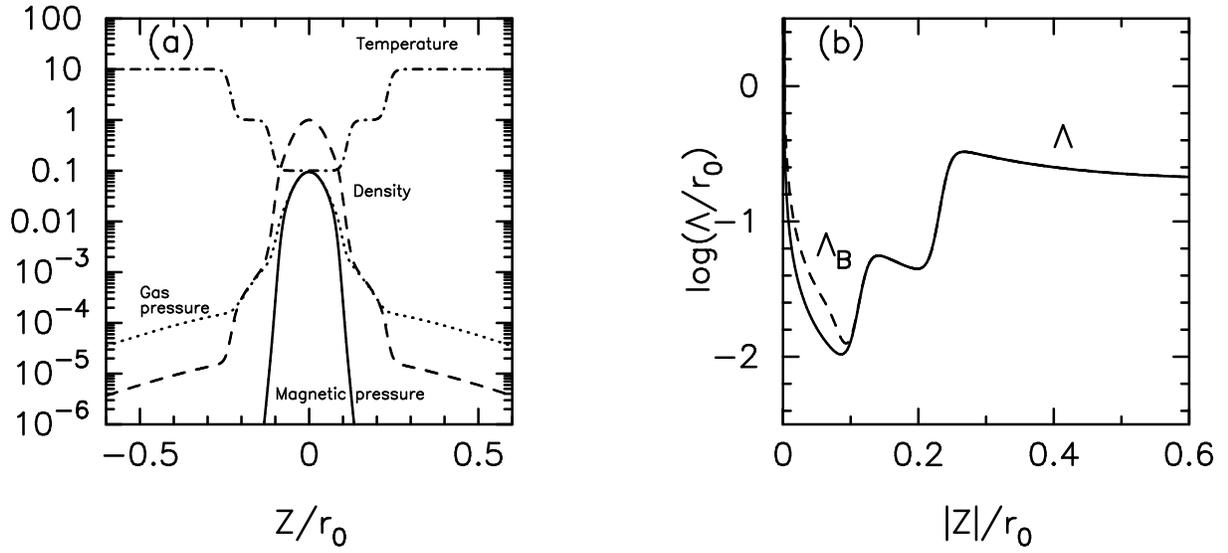}
\end{center}
\caption{(a) Distribution of the initial density (dashed curve), gas pressure (dotted curve),
magnetic pressure (solid curve), and temperature (dash dotted curve), for the fiducial model (model B).
(b) Distribution of the initial local pressure scale height $\Lambda$ (solid curve) and modified scale
height $\Lambda_\mathrm{B}=(1+\beta(z)^{-1})\Lambda$ (dashed curve) for the fiducial model (model B).}
\end{figure}

\begin{figure}
\begin{center}
\FigureFile(80mm,80mm){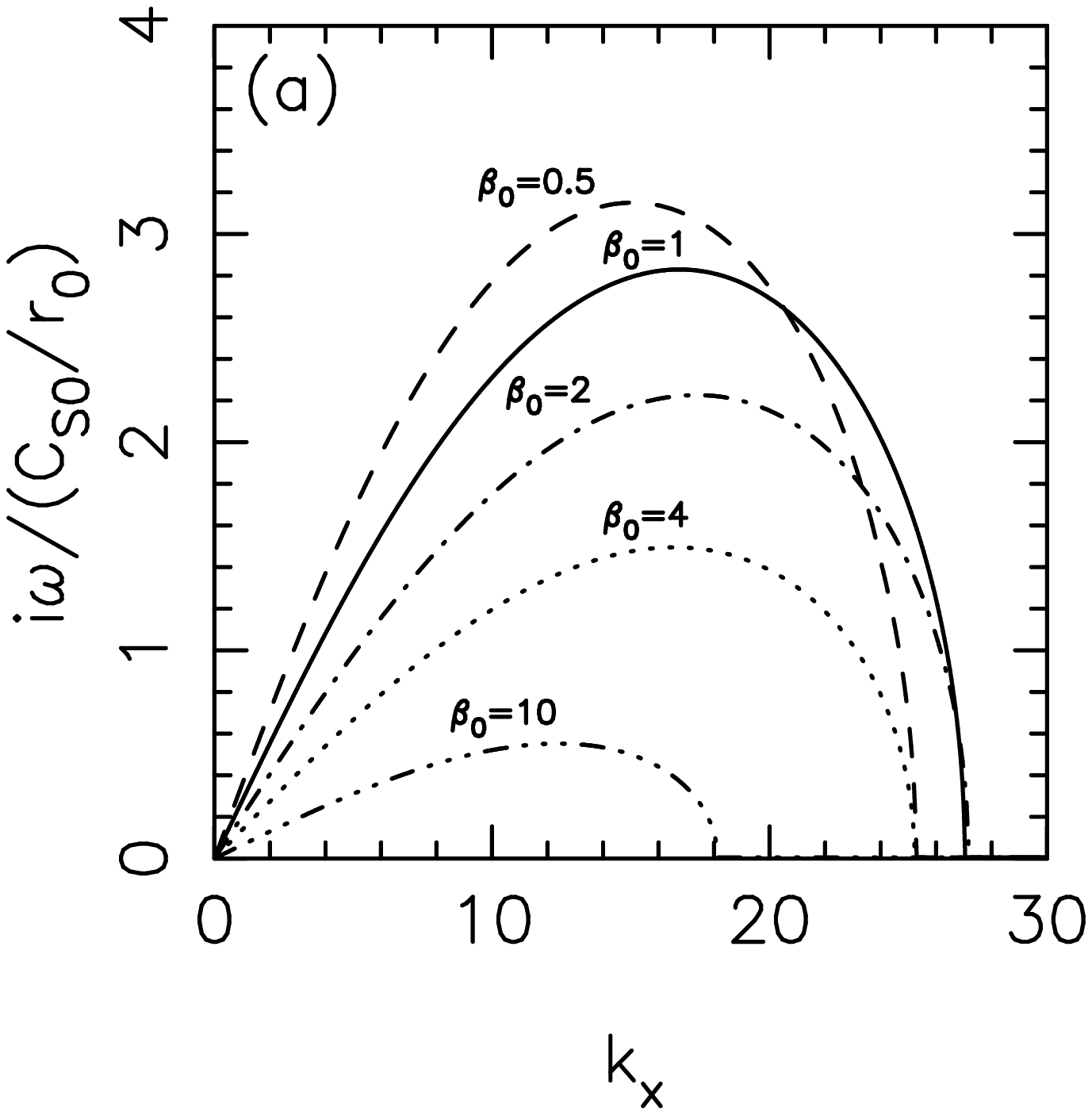}
\FigureFile(80mm,80mm){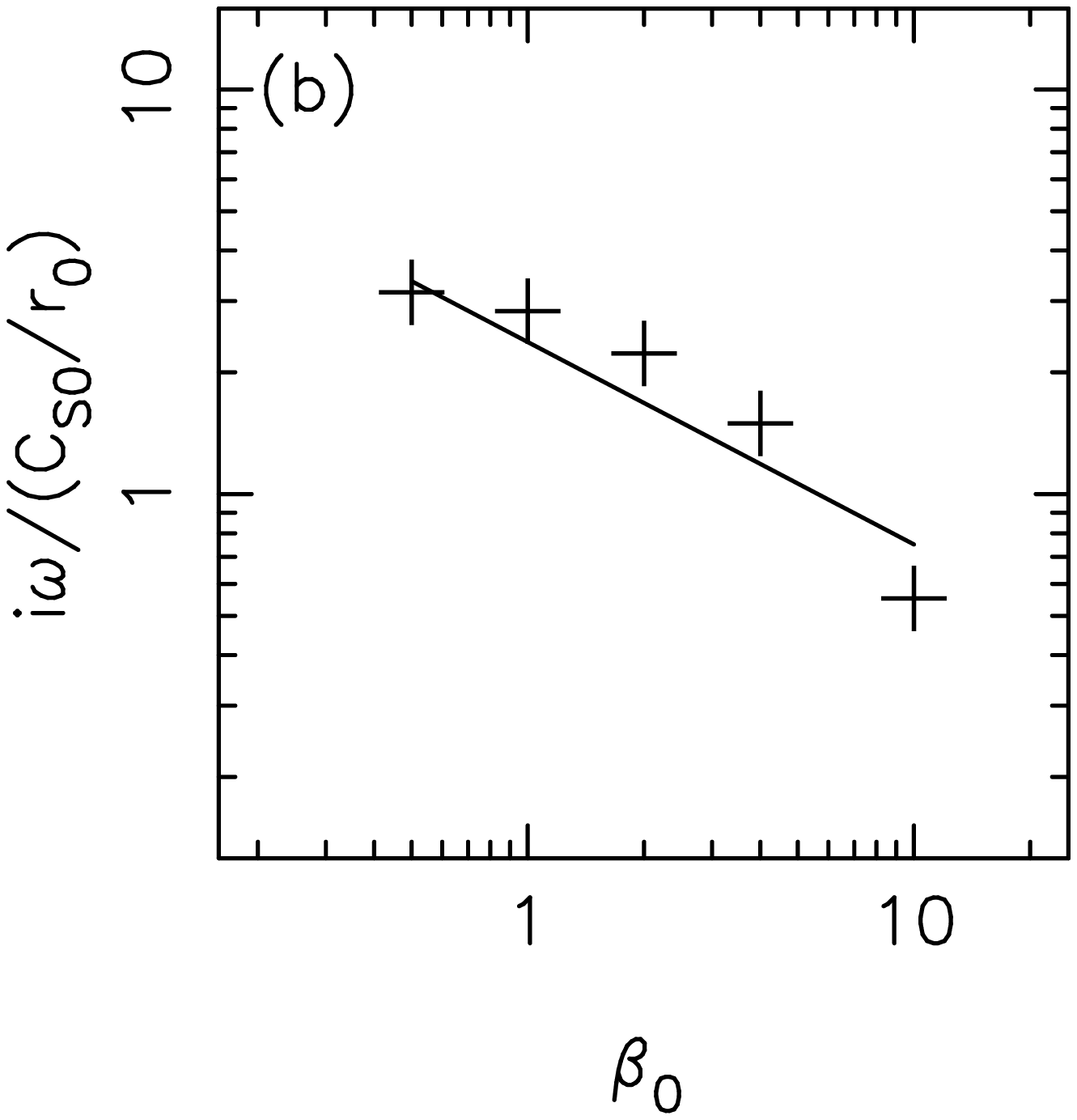}
\end{center}
\caption{(a) Linear growth rate of the Parker instability for the unperturbed states with $\beta_0=0.5,1,2,4$ and
10 as a function of wave number $k_\mathrm{x}$. (b) The dependence of the maximum growth rates on the initial plasma beta.
The solid line shows a line where $i\omega \propto \beta_0^{-1/2}$. The unit of the growth rate is $C_\mathrm{S0}/r_0=1/t_0$.}
\end{figure}

\begin{figure}
\begin{center}
\FigureFile(160mm,240mm){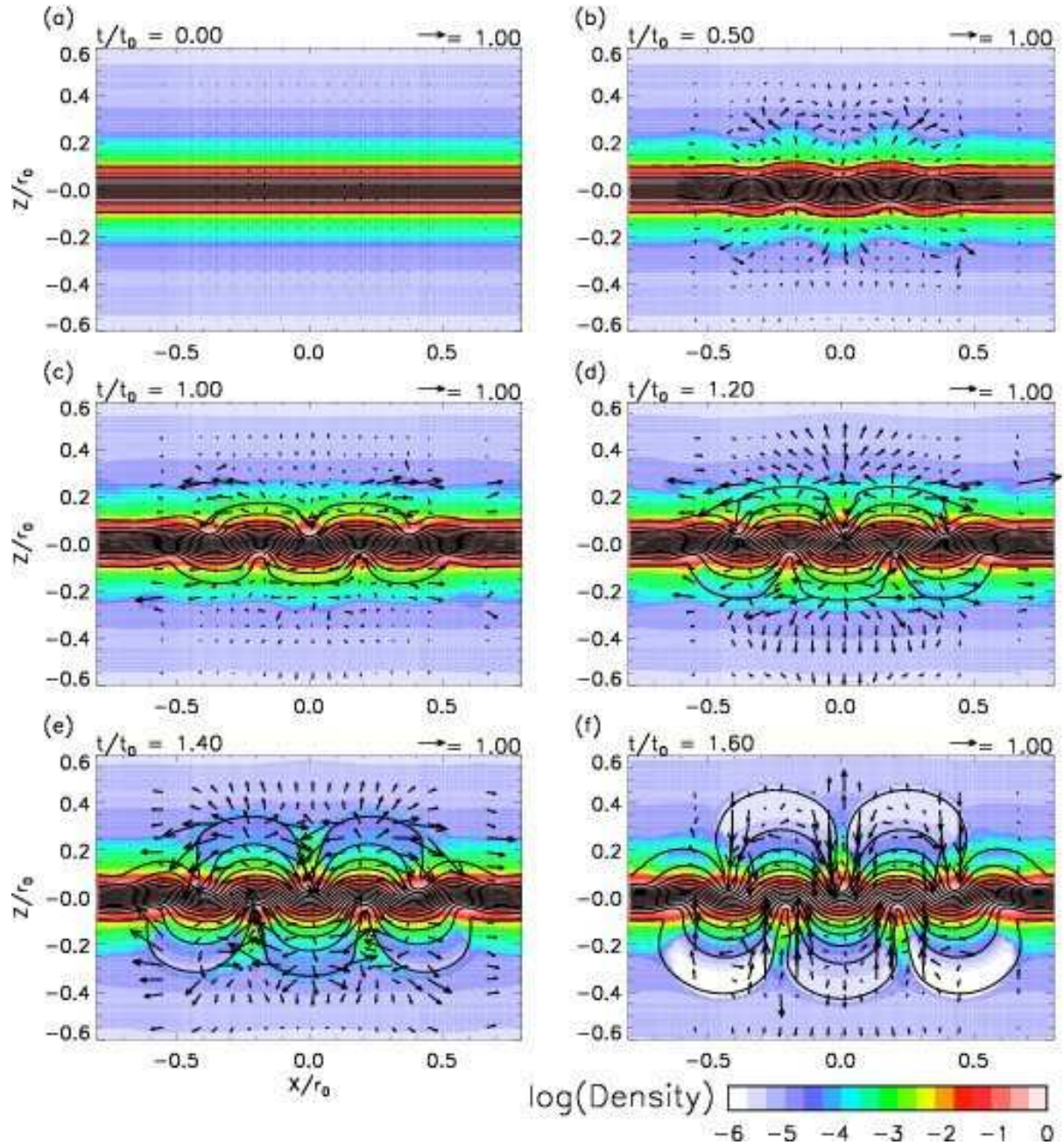}
\end{center}
\caption{Numerical results for the fiducial model (model B); density in a logarthmic scale (colors), magnetic field
lines (solid curves), and velocity field (vectors) for $t/t_0=0.0,0.5,1.0,1.2,1.4$ and 1.6.}
\end{figure}

\begin{figure}
\begin{center}
\FigureFile(80mm,160mm){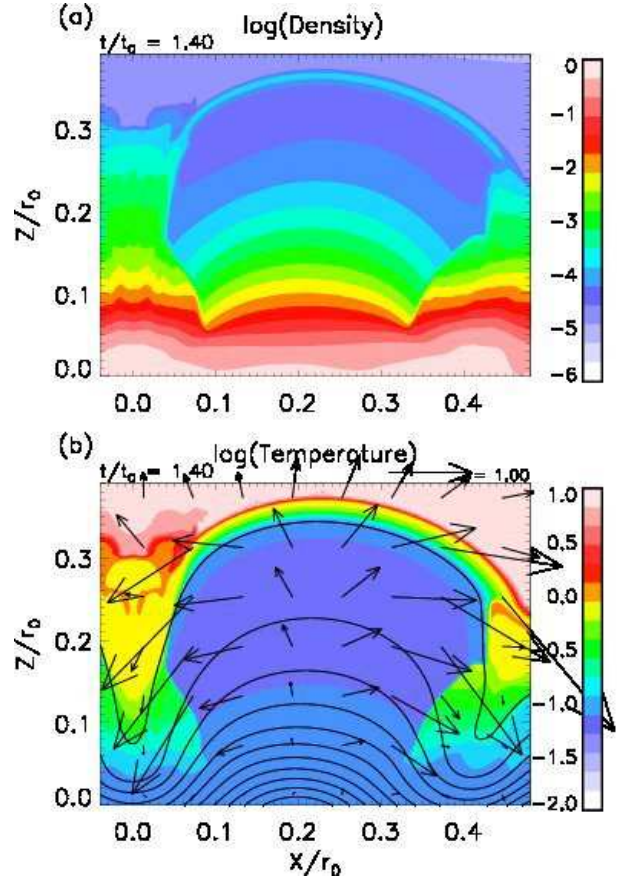}
\end{center}
\caption{Numerical result for the fiducial model (model B; $\beta_0=1$); (a) density in a logarithmic scale (colors)
at $t/t_0=1.4$. (b) temperature in a logarthmic scale (colors), magnetic field lines (solid curves), and velocity
field (vectors) at $t/t_0=1.4$.}
\end{figure}

\begin{figure}
\begin{center}
\FigureFile(80mm,80mm){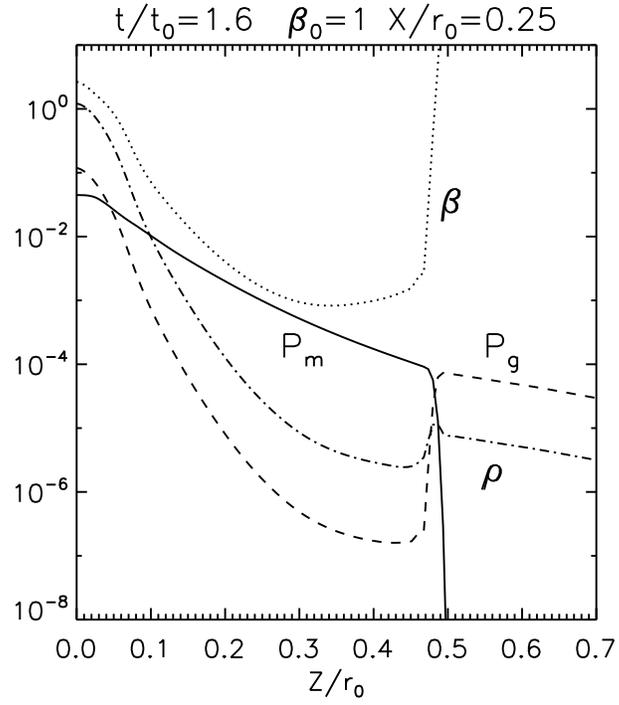}
\end{center}
\caption{One-dimensional ($z$-)distribution of the logarithmic magnetic pressure $P_\mathrm{m}$ 
(solid curve), gas pressure $P_\mathrm{g}$ (broken curve), plasma beta $\beta$ (dotted curve) 
and density $\rho$ (dash dotted curve) at $t/t_0=1.6$ for the fiducial model (model B).}
\end{figure}

\begin{figure}
\begin{center}
\FigureFile(160mm,240mm){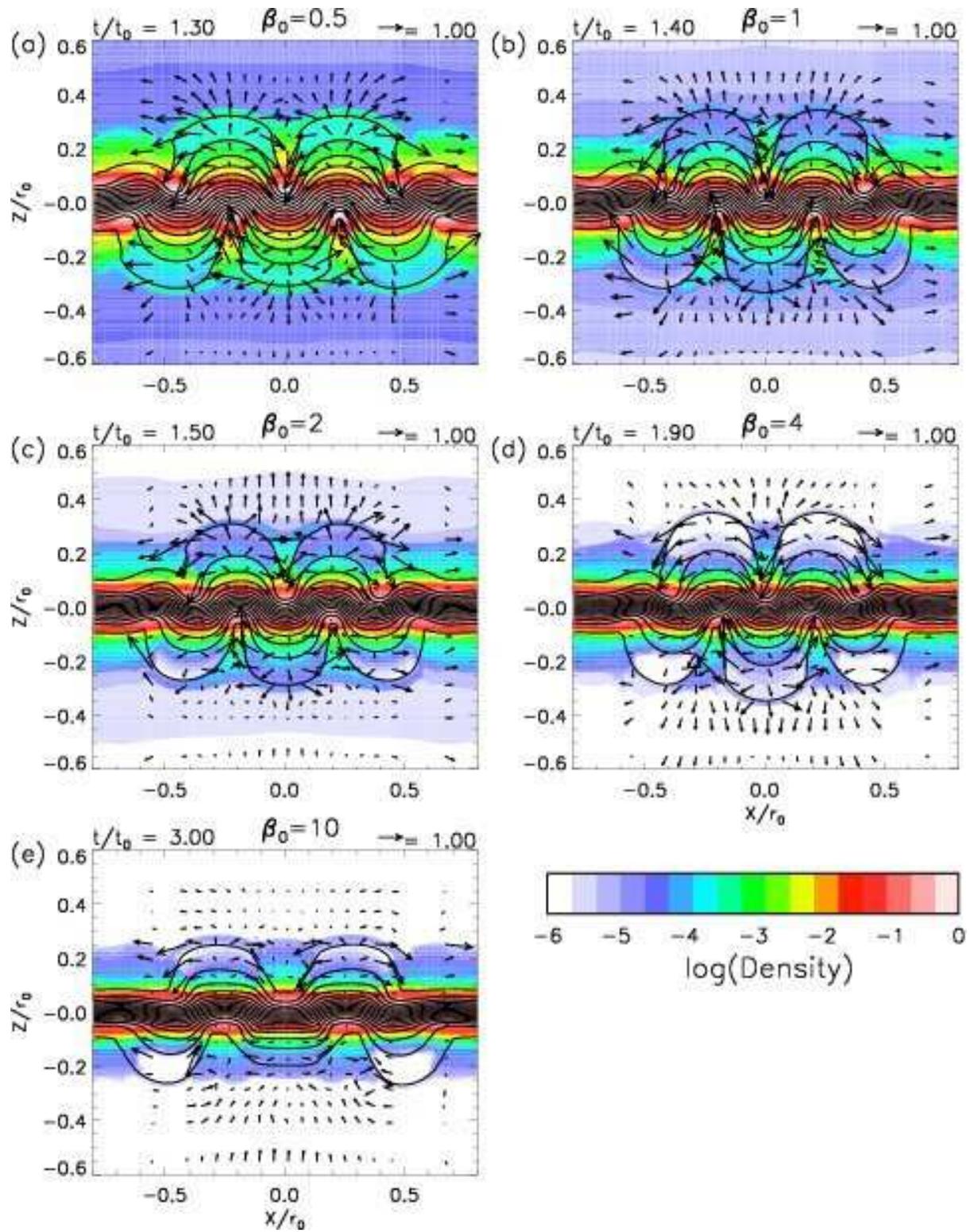}
\end{center}
\caption{Numerical results for all models ($\beta_0=0.5,1,2,4$ and 10); density in a logarthmic scale (colors), magnetic
field lines (solid curves), and velocity field vectors.}
\end{figure}

\begin{figure}
\begin{center}
\FigureFile(80mm,120mm){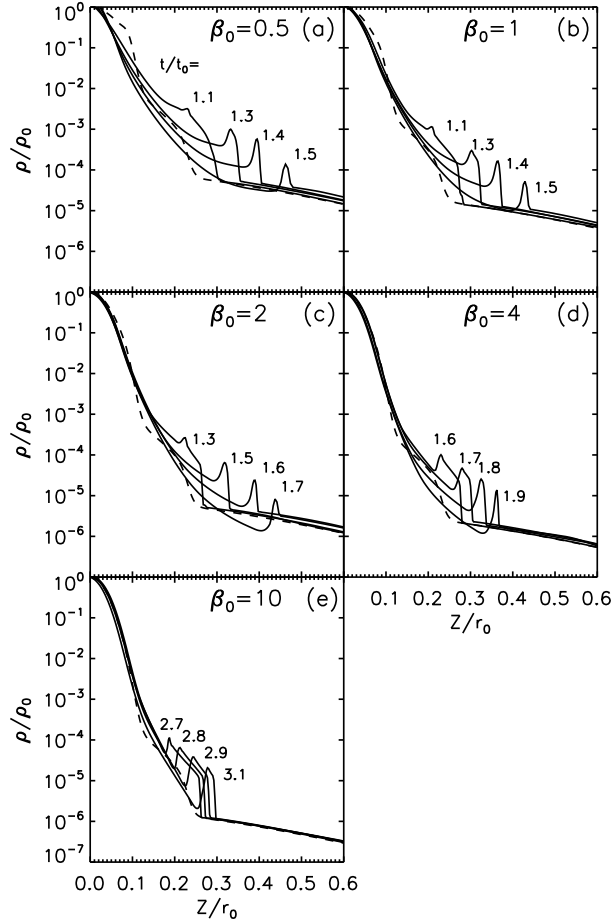}
\end{center}
\caption{Distribution of the logarithmic density in the $z$-direction measured at the loop center
for models (a) model A, (b) model B, (c) model C, (d) model D and (e) model E at various times.}
\end{figure}

\begin{figure}
\begin{center}
\FigureFile(80mm,80mm){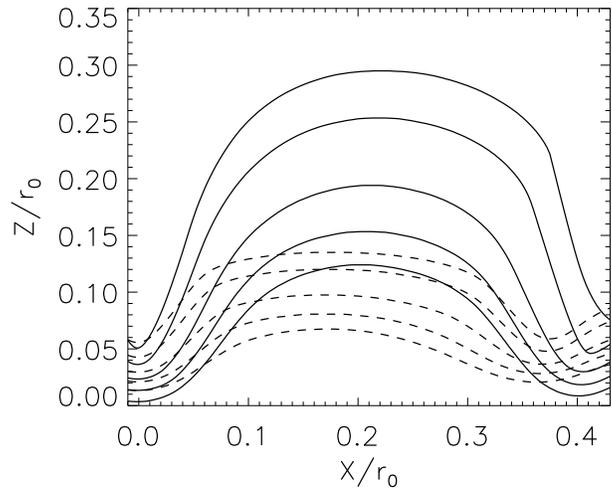}
\end{center}
\caption{Magnetic field lines at $t/t_0=1.4$ (solid curves) and at $t/t_0=0.9$ (broken curves) for the fiducial model.}
\end{figure}

\begin{figure}
\begin{center}
\FigureFile(160mm,240mm){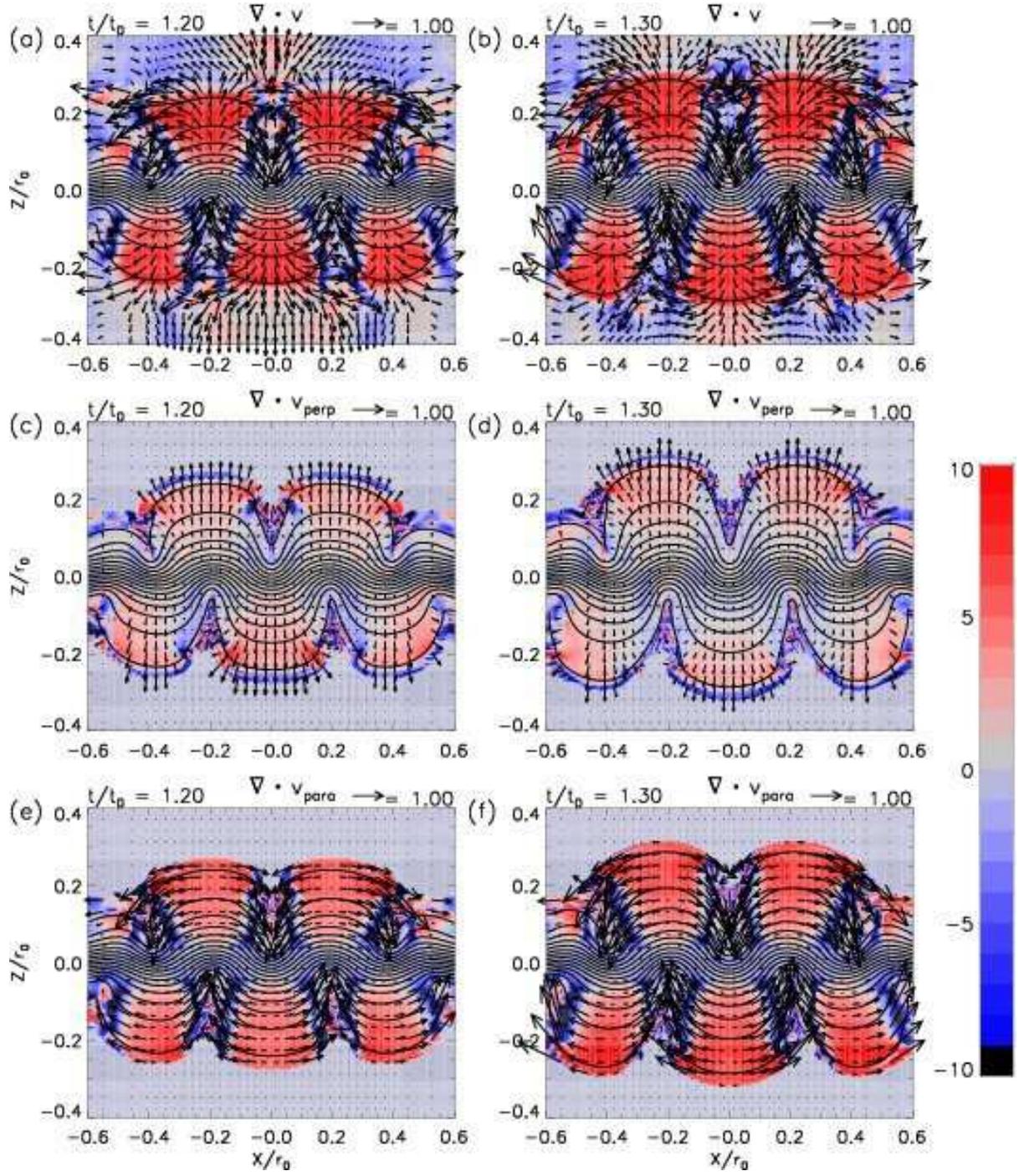}
\end{center}
\caption{(a,b) Divergence of velocity at $t/t_0 = 1.2$ and 1.3. Solid curves show magnetic field lines, and arrows
show the velocity vector, (c,d) for $\boldsymbol{v}_\perp$, and (e,f) for $\boldsymbol{v}_\parallel$. Colors in figures
(c,d) and (e,f) show the divergence of velocity components perpendicular and parallel to the magnetic field, respectively.}
\end{figure}

\begin{figure}
\begin{center}
\FigureFile(160mm,80mm){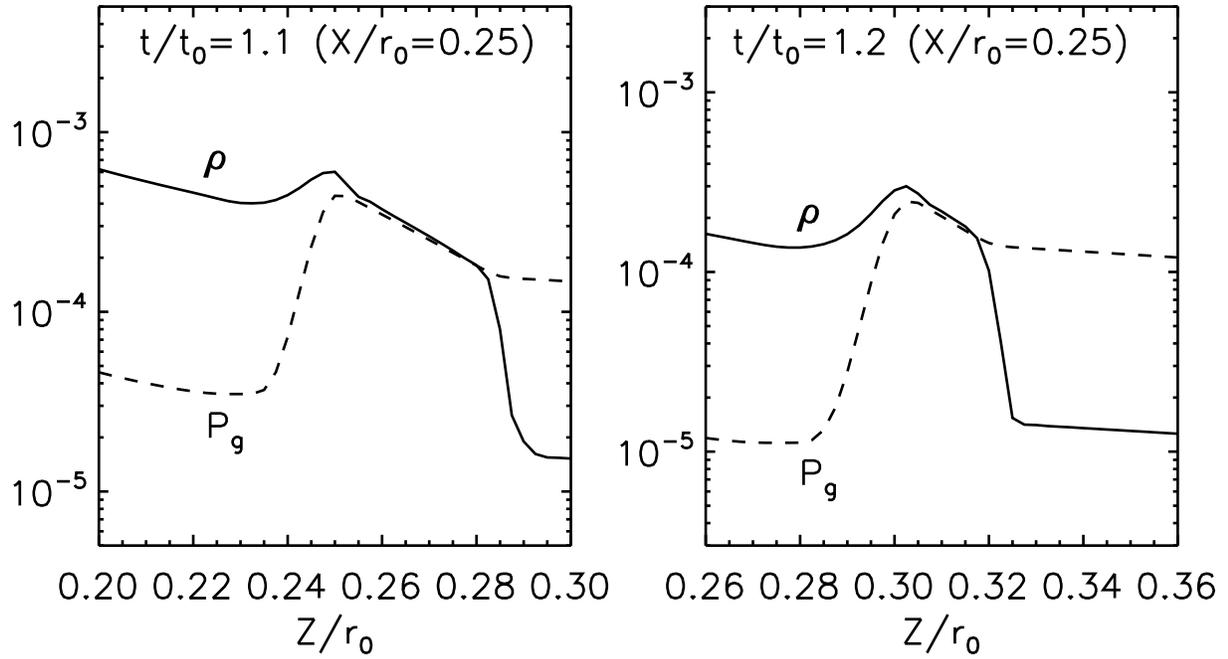}
\end{center}
\caption{Distribution of the logarithmic density (solid curve) and pressure (broken curve) near the loop top at $x/r_0=0.25$
for the fiducial model (model B) at $t/t_0=1.1$ (left) and $t/t_0=1.2$ (right).}
\end{figure}

\begin{figure}
\begin{center}
\FigureFile(160mm,160mm){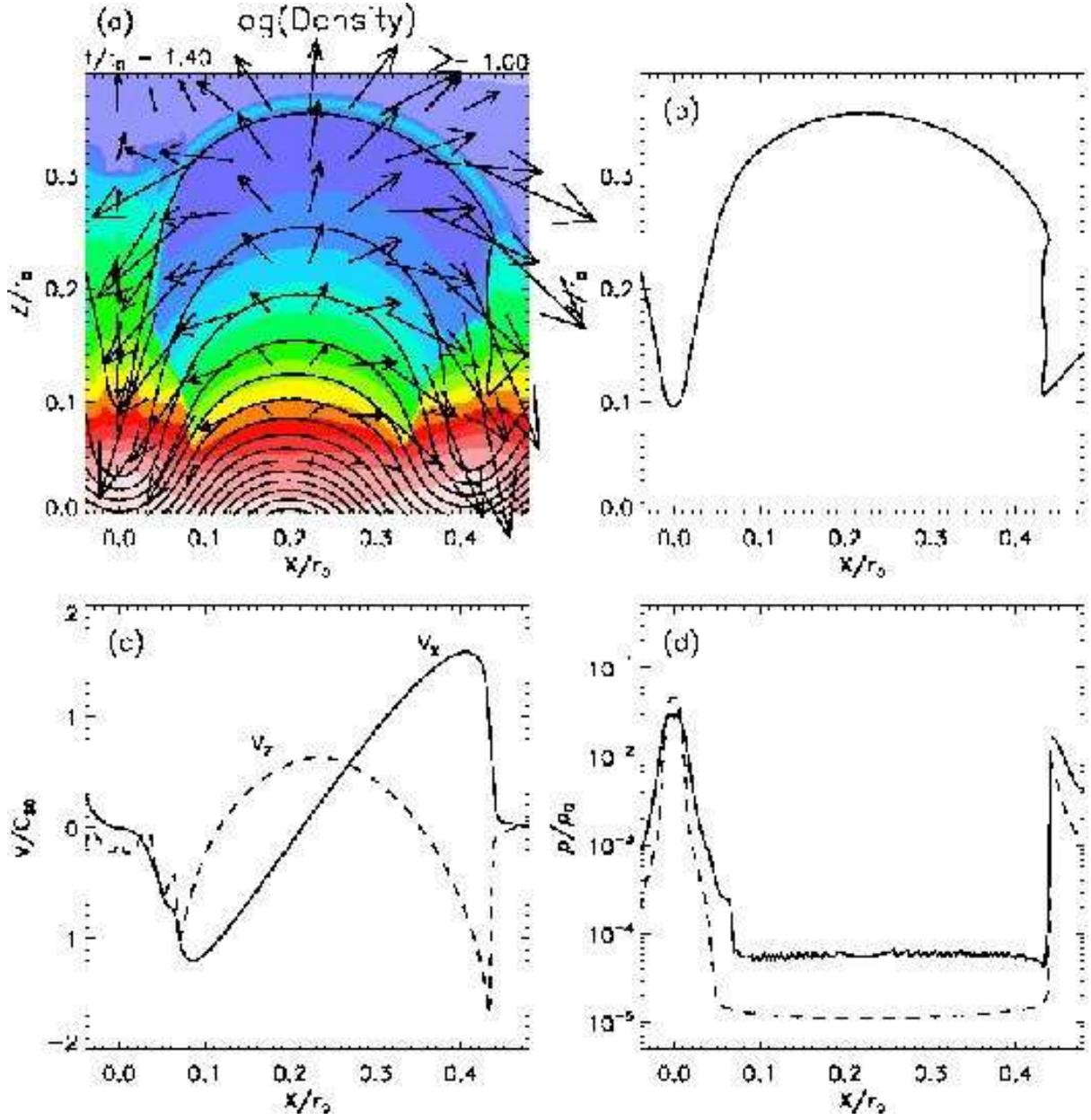}
\end{center}
\caption{(a) Numerical results for the fiducial model (model B); density in a logarthmic scale (colors),
magnetic field lines (solid curves), and velocity field vectors at $t/t_0=1.4$. (b) The outermost magnetic
field line at $t/t_0=1.4$. (c) Velocity components ($v_x, v_z$) along the outermost magnetic field line.
(d) Density along the outermost magnetic field line at $t/t_0=1.4$ (solid curve) and $t/t_0=0$ (broken curve). }
\end{figure}

\begin{figure}
\begin{center}
\FigureFile(100mm,100mm){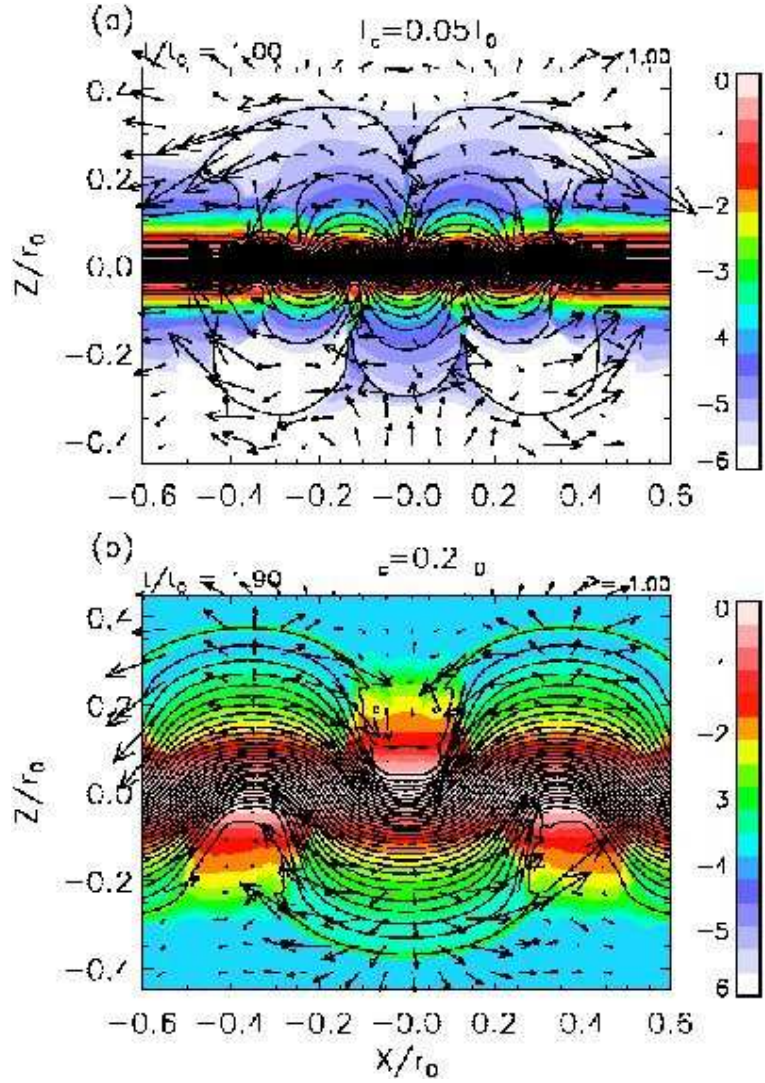}
\end{center}
\caption{Dependence of the numerical results on the initial  disk temperature. Color shows the density distribution
at the stage when the top of the magnetic loops reaches $z/r_0 \sim 0.35$.  (a) $T_\mathrm{c}=0.05T_0$ at $t/t_0=1.0$.
(b) $T_\mathrm{c}=0.2T_0$ at $t/t_0=2.0$. Solid curves show magnetic field lines. Arrows show velocity vectors.}
\end{figure}

\end{document}